\documentclass{iaus}
\usepackage{graphicx}

\title[Magnetic fields, winds and X-rays of massive stars] 
{Magnetic fields, winds and X-rays of massive stars in the Orion Nebula Cluster}

\author[V. Petit, G. A. Wade, L. Drissen, T. Montmerle, E. Alecian]   
{V. Petit$^1$, G. A. Wade$^2$, L. Drissen$^1$, T. Montmerle$^3$ \and E. Alecian$^2$}

\affiliation{$^1$D\'epartement de Physique, Centre de recherche en Astrophysique du Qu\'ebec, Universit\'e Laval, Qu\'ebec, Canada \break $^2$Department Physics, Royal Military College of Canada, Kingston, Canada, \break $^3$Laboratoire d'Astrophysique de Grenoble, Universit\'e Joseph Fourier,  CNRS, Grenoble, France}

\pubyear{2009}
\volume{259}  
\pagerange{100--100}
\date{?? and in revised form ??}
\jname{Cosmic Magnetic Fields: From Planets, to Stars and Galaxies}
\editors{K.G. Strassmeier, A.G. Kosovichev \& J.E. Beckman, eds.}
\begin{document}

\maketitle

\begin{abstract}
In massive stars, magnetic fields are thought to confine the outflowing radiatively-driven wind, resulting in X-ray emission that is harder, more variable and more efficient than that produced by instability-generated shocks in non-magnetic winds. Although magnetic confinement of stellar winds has been shown to strongly modify the mass-loss and X-ray characteristics of massive OB stars, we lack a detailed understanding of the complex processes responsible. The aim of this study is to examine the relationship between magnetism, stellar winds and X-ray emission of OB stars. In conjunction with a Chandra survey of the Orion Nebula Cluster, we carried out spectropolarimatric ESPaDOnS observations to determine the magnetic properties of massive OB stars of this cluster.
\end{abstract}

\begin{figure}
\centering
 \includegraphics[width=9.5cm]{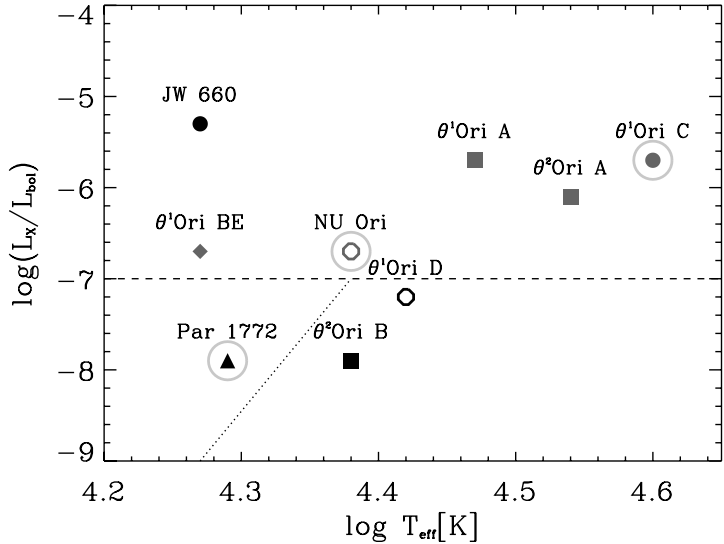}
  \caption{X-ray efficiency of the ONC massive stars as a function of effective temperature (Stelzer et al. 2005). The 3 detected stars are circled. Filled symbols are for stars with indirect indications of the presence of a magnetic field, and grey symbols are for confirmed or suspected binaries. Plotting symbols indicate the following properties: circles are for stars showing possible X-ray rotational modulation, squares are for T\,Tauri type X-ray emission, triangles are chemically peculiar (CP) stars, and the diamond star was not observed. The dotted lines indicates the typical efficiency for massive O-type stars, and a qualitative illustration of the sharp decrease in the B-type stars range.}\label{fig:fig1}
\end{figure}

Magnetic fields are well known to produce X-rays in late-type convective stars like the Sun.
On the other hand, X-ray emission from massive stars is thought to have a different origin. Their powerful winds are driven by by radiation pressure through spectral lines, which is an unstable process. These instabilities result in collisions between wind streams of different velocity, resulting in small shocks that generate X-ray emission (\cite{1999ApJ...520..833O, 1980ApJ...241..300L}).

The Chandra Orion Ultradeep Project (COUP) was dedicated to observe the Orion Nebula Cluster (ONC) in X-rays. The sample of 20 O, B and A-type stars was studied with the goal of disentangling the respective roles of winds and magnetic fields in producing X-rays (\cite{2005ApJS..160..557S}).
The production of X-rays by radiative shocks should be the dominant mechanism for the subsample of 9 O to early-B stars. However, aside from 2 of those stars, all targets showed X-ray intensity and/or variability which were inconsistent with the small shock model predictions (Figure 1).

To examine the effect of magnetic fields on the winds of the massive stars in the ONC, we conducted spectropolarimetric observations in order to directly detect and characterise potential magnetic fields in these stars, through the Stokes V circular polarisation induced by the Zeeman effect. 
We used the ESPaDOnS spectropolarimeter at CFHT in 2006 and 2007 to observe 8 of the 9 ONC massive stars. Additional measurements of $\theta^1$\,Ori\,C and HD\,36982 were taken with ESPaDOnS in December 2007 and with ESPaDOnS's twin Narval, installed at T\'elescope Bernard Lyot, in November 2007. 

From our observations, we found clear Stokes V magnetic signatures for three stars: the previously-detected $\theta^1$\,Ori\,C, as well as HD\,36982 and HD\,37061 (\cite{2008MNRAS.387L..23P}). The magnetic stars are encircled in Figure 1. 
The Stokes V signatures were directly modelled with the polarised LTE radiative transfer code \textsc{zeeman2} (\cite{1988ApJ...326..967L}), leading to an inferred surface dipolar field strength of $1150^{+320}_{-200}$\,G and $620^{+220}_{-170}$\,G for HD\,39682 and HD\,37061 respectively. 

This study of the Orion stellar cluster represents a complete magnetic survey of a co-evolved and co-environmental population of massive stars. 
The 3 magnetic stars of the ONC have fields that should be strong enough to dynamically influence their stellar winds at a significant level. However, this field-wind interaction is not reflected in any systematic way in the X-ray properties of these stars. Furthermore, no fields are found (with typical upper limits on the surface dipole strength of the order of 100\,G) in other ONC massive stars that Stelzer et al. (2005) considered to be ``prime candidates'' for magnetism. From this we conclude that: (i) X-ray variability, intensity and hardness enhancement are not systematically correlated with the presence of a magnetic field. More detailed studies of the field geometries of these magnetic stars will serve as inputs to new models (\cite{2007MNRAS.382..139T}) and 3D MHD simulations of magnetic wind confinement (\cite{2008MNRAS.385...97U}), to better understand the mechanisms that lead to this variety of X-ray properties. 
(ii) The classical Lucy \& White (1980) mechanism for soft X-ray production in non-magnetic winds requires significant 
revision to explain the characteristics of some stars in the ONC.

\end{document}